\documentstyle[twoside,fleqn,espcrc2]{article}

\font\mybb=msbm10 at 10pt
\def\bb#1{\hbox{\mybb#1}}
\def\R {\bb{R}}
\def\Z {\bb{Z}}

\newcommand{\AmS}{{\protect\the\textfont2
  A\kern-.1667em\lower.5ex\hbox{M}\kern-.125emS}}

\title{Supergravity solitons and non-perturbative superstrings}

\author{P.K. Townsend\address{DAMTP, University of Cambridge,\\
       Silver St., Cambridge, U.K.}}

\begin{document}

\begin{abstract}
A review is given of the implications of supersymmetric black holes for the
non-perturbative formulation of toroidally compactified superstrings, with
particular emphasis on symmetry enhancement at special vacua and
S-duality of the heterotic string.
\end{abstract}

\maketitle

Heterotic and type II superstrings have, when toroidally-compactified to D=4,
N=4 and N=8 supersymmetry, respectively. The (generic) effective N=4 or N=8
supergravity theory has, in both cases, 28 U(1) gauge fields and hence
particles in
the spectrum may carry any of 56 electric or magnetic charges; we may therefore
associate to each particle in the spectrum a charge 56-vector ${\cal Z}$. In
perturbation theory only the 28 electric charges are carried by particles in
the
heterotic string spectrum, and only 12 of these occur in the perturbative type
II superstring spectrum. The remaining charges are carried by particles with
masses that go to infinity as the string coupling constant $g$ goes to zero so
they are absent in the free string spectrum, but they can be found by a
semi-classical analysis based on the effective N=4 or N=8 supergravity theory.

Recent advances in our understanding of the non-perturbative dynamics of these
theories have made essential use of the existence of this non-perturbative
spectrum, with the main analytical tool being the Bogomol'nyi bound
\cite{Bog76,WO78,GH82} on the the mass of any particle in terms of its charge
vector. All particles in the free string spectrum are, of course, found to
satisfy the bound, but few of them saturate it. However, when $g\ne0$,
particles
that do {\it not} saturate the bound can be expected to become unstable against
decay into particles which do saturate it. For example, massive particles in
the
perturbative string spectrum which fail to saturate the bound by a large margin
can be viewed as (non-extreme) black holes which will decay by Hawking
radiation. It therefore seems reasonable to suppose that, at non-zero string
coupling, the only stable states are those which saturate the Bogomol'nyi
bound.
If this is the case, then every stable particle in the spectrum has a mass
given
by the formula
\begin{equation}
M^2 = {\cal Z}^T {\cal R} (\langle\phi\rangle) {\cal Z}\ .
\label{eq:onea}
\end{equation}
where ${\cal R}$ is a $56\times 56$ matrix that depends on the expectation
values $\langle\phi\rangle$ of the massless scalar fields $\phi$ of the
effective supergravity theory, i.e. on the choice of vacuum.

In the N=8 case, the space of vacua parametrized by the scalar field
expectation
values $\langle\phi\rangle$ is locally isomorphic to the coset space
$E_{7,7}/SU(8)$ \cite{CJ}, and the bound (\ref{eq:onea}) is invariant
under the natural action of $E_{7,7}$ on this space if ${\cal Z}$ is taken to
transform as the ${\bf 56}$ of $E_{7,7}$. U-duality of the type II superstring
theory \cite{HT95a} implies that the actual space of vacua is
\begin{equation}
E_7(\Z)\backslash E_{7,7}/SU(8)
\label{eq:aonea}
\end{equation}
where $E_7(\Z)$ is the discrete U-duality subgroup of $E_{7,7}$ that contains
the $S\times T$ duality group $Sl(2;\Z)\times SO(6,6;\Z)$ of the type II
superstring. The matrix ${\cal R}$ appearing in the N=8
Bogomol'nyi bound is non-singular, so all charged particles have non-zero
masses. This is a consequence of the fact that {\it all} 56 charges occur as
central charges in the N=8 supersymmetry algebra and is clearly necessary for
the complete symmetry between all 56 types of charge implied by U-duality.

In the heterotic case, there are only 12 central charges in the N=4
supersymmetry
algebra, so there must be a 44-dimensional subspace of the full 56-dimensional
charge vector space for which the bound implies merely that the mass
$M$ is non-negative. Indeed, in the heterotic case the matrix ${\cal R}$ has 44
zero eigenvalues for all scalar field expectation values \cite{HT95b}. To
investigate the nature of the zero-eigenvalue eigenspace it is convenient to
split the scalar field expectation values, which we have denoted generically by
$\langle\phi\rangle$, into the complex axion/dilaton expectation value $\lambda
= \theta/2\pi +i/g^2$, where $g$ is the string coupling constant and $\theta$ a
vacuum angle, and the remaining moduli $\varphi$, which take values in the
coset
space $SO(6,22)/[SO(6)\times SO(22)]$. We can then set ${\cal Z}^T =(p,q)$,
where $p$ denotes the magnetic charges and $q$ the electric ones, and rewrite
the mass formula (\ref{eq:onea}) in the form \cite{SS94}
\begin{equation}
M^2 =  \pmatrix {p & q} [{\cal S}(\lambda) \otimes {\cal
T}(\varphi)]  \pmatrix{p \cr q}\ ,
\label{eq:oneb}
\end{equation}
where the matrix ${\cal T}$ acts on the electric and magnetic 28-vectors $q$
and $p$, and the $2\times 2$ matrix ${\cal S}$ acts on the 2-vector $(p,q)$.
This mass formula is invariant under the action of
\begin{equation}
Sl(2;\R)\times SO(6,22)
\label{eq:onec}
\end{equation}
on the charge 56-vector $(p,q)$ provided that the moduli $\lambda$ and
$\varphi$
are simultaneously transformed. We note that T-duality
of the heterotic string theory implies that the space parameterized by the
moduli $\varphi$ is actually
\begin{equation}
SO(6,22;\Z)\backslash SO(6,22)/[SO(6)\times SO(22)]\ ,
\label{eq:oned}
\end{equation}
i.e. vacua which differ by an $SO(6,22;\Z)$ transformation are equivalent.
Similarly, S-duality \cite{FILQ90} implies that the space parameterised by
$\lambda$ is actually
\begin{equation}
Sl(2;\Z)\backslash Sl(2;\R)/U(1)\ .
\label{eq:aoned}
\end{equation}
Unlike S-duality, which is invisible in perturbation theory, T-duality has
consequences for the perturbative heterotic or type II superstring. These
predictions of T-duality have been verified (see \cite{Trev} for a review) but
this is  insufficient to establish T-duality as a property of the full
non-perturbative superstring theory; at this level S,T and U duality are {\it
all} conjectural, although evidence in their support continues to mount.

The matrix ${\cal S}$ in (\ref{eq:oneb}) is non-singular for all values of
$\lambda$. The reason that the $56\times 56$ matrix ${\cal R}$ has 44 zero
eigenvalues is that the $28\times 28$ matrix ${\cal T}$ has 22 zero
eigenvalues.  Let us  concentrate on the electric charge 28-vectors $q$. They
can be classified according to whether their $SO(6,22)$ norm $q^2$ is positive,
zero or negative. The 22-dimensional kernel of ${\cal T}$ is spanned by {\it
negative} norm 28-vectors, but a negative norm is only a necessary, and not a
sufficient, condition for a vector to lie in the kernel of ${\cal T}$. Of
course,
given a negative-norm charge vector that is not in this kernel, there is an
$SO(6,22)$ transformation that will take it into the kernel. However, this
transformation will also change the vacuum. Thus, for a given vacuum, a purely
electric particle with a {\it given} negative norm electric charge vector will
not generally be massless, but there will be vacua in which it {\it is}
massless \cite{HT95b}. Instead of changing the vacuum we
could just start with a different negative norm charge vector (the possible
choices differ by $SO(6,22)$ transformations) and it might appear from this
that
charged massless  particles must occur for {\it any} choice of vacuum. But we
have yet to take into account the charge quantization required by quantum
mechanics.

Charge quantization requires, for any given vacuum, that the charge vector lie
in a lattice. An $SO(6,22)$ transformation of a charge vector (for fixed
vacuum) will generally take a vector in this lattice to one that is not in the
lattice. In fact, it will do so unless the $SO(6,22)$ transformation is
actually
an $SO(6,22;\Z)$ transformation, where the embedding of the discrete subgroup
$SO(6,22;\Z)$ in $SO(6,22;\Z)$ depends on the vacuum. Thus, the allowed choices
of initial negative norm charge vectors belong to a discrete, rather than a
continuous, set. This set is the union of $SO(6,22;\Z)$ orbits of lattice
vectors, each orbit being characterised by an $SO(6,22)$ norm.

Generically,
none of these allowed charge vectors will be in the kernel of ${\cal T}$ but,
as explained above, there will be vacua in which any one of them is. If there
were particles in the spectrum with arbitrarily large negative $q^2$, then
these
special vacua would be dense in the space of all vacua. It is difficult to see
how one could make physical sense of this. Fortunately, this doesn't happen
because the `Bogomol'nyi' states in the perturbative heterotic string spectrum
all have $q^2=-2 +2N_L$ where the integer $N_l$ is the oscillator number of the
left  moving worldsheet fields (see, for example, \cite{S94}). Since $N_L$ is
non-negative, the only perturbative states with negative $q^2$ are those for
which $N_L=0$, which have $q^2=-2$. In this context, the fact that there
exist special vacua in which these states are massless is well-known as the
Halpern-Frenkel-Ka{\v c} mechanism.

It seems possible that this restriction on
the allowed negative values of $q^2$ may have a general explanation (as just
remarked, the absence of any restriction leads to paradoxical conclusions) but
this
explanation is lacking at present. Since we have invoked results already
established
in perturbation theory to restrict the possible charge vectors, it might seem
that
little has been gained from the prior general analysis based on the Bogomol'nyi
mass formula. However, the mass formula does give additional information
because
it tells us that a vacuum in which electrically charged particles are massless
is also a vacuum in which magnetically charged particles are massless
\cite{HT95b}. In fact, if S-duality is assumed, each additional massless
particle will be one member of an entire $Sl(2;\Z)$ orbit of additional
massless
particles.

This explains an otherwise puzzling feature of the HFK mechanism. T-duality
requires that the massless effective field theory have an $SO(6,22)$ invariance
of the equations of motion. The perturbative HFK mechanism suggests that the
effective field theory in a vacuum with enhanced symmetry is a locally
supersymmetric, rank 28, N=4 {\it non}-abelian gauge theory, but such theories
are {\it not} $SO(6,22)$ invariant. The resolution is that the evidence of
perturbation theory is misleading because the fact that magnetically charged
particles must become massless simultaneously with the electric ones means that
the physics in these enhanced symmetry vacua cannot be described by a local
field theory.

Another obvious advantage of the approach to symmetry enhancement via the
Bogomol'nyi mass formula is that it depends only on N=4 supersymmetry and
therefore applies equally to the $K_3\times T^2$ compactified type II
superstring \cite{HT95b}, thus providing evidence for the proposal \cite{HT95a}
that these two string theories are non-perturbatively equivalent.

That the particles in the spectrum with negative $q^2$ are the ones
responsible for enhanced symmetry at special vacua is not surprising when
one considers that by switching off the supergravitational interactions we
remove the six gauge fields in the graviton multiplet and break the
$SO(6,22)$ invariance to $SO(22)$. We then have an N=4 $U(1)^{22}$ gauge theory
coupled to massive charged N=4 supermultiplets for which the charge vectors
automatically have negative norm. These multiplets have a natural
interpretation as Higgs supermultiplets. Their magnetic duals are BPS
monopoles \cite{S94}. In this `Higgs' sector the conjectured S-duality of the
heterotic string can therefore be viewed as generalization of Montonen-Olive
duality \cite{MO77,GNO77}.

The connection between field theory solitons and particles in the spectrum of
the quantum theory is usually made via semi-classical quantization. These
methods are generally reliable at weak coupling, but one could question their
applicability near those special vacua of the heterotic string at which
otherwise massive solitons become massless. Indeed, one might well expect the
whole soliton picture to break down at these points. It is therefore rather
surprising to learn that this does not happen, in the sense that {\it massless}
`soliton' solutions can be found \cite{KL95,CY95}. They are singular, however,
and their significance for the quantum theory has still to be elucidated.

We now move on to the electric states with $q^2\ge0$. Since only the
$q^2=-2$ states involve the (broken) non-abelian group structure one might
suspect that these should appear as solutions of the effective
massless $U(1)^{28}$ supergravity theory. To explore this possibility it is
convenient to consider the consistent truncation of the effective supergravity
theory of the heterotic string to one with the Lagrangian \begin{equation}
L= \sqrt{-g}\Big[ R -2(\partial\sigma)^2 - e^{-2a\sigma}F^2\Big]
\label{eq:twoa}
\end{equation}
where $F=dA$ is a Maxwell field-strength two-form and $\sigma$ is a scalar
field which is some function of the scalars $\phi$; i.e. some combination
of the dilaton and the other moduli fields. We may assume without
loss of generality that $a$ is positive since the field redefinition
$\sigma\rightarrow -\sigma$ effectively changes the sign of $a$. The only
values
of $a$ that arise in this way are \cite{HT95a}
\begin{equation}
a= 0,{1\over\sqrt{3}},1,\sqrt{3}\ .
\label{eq:twob}
\end{equation}
Strictly speaking, the value $a=0$ does not arise in this way, but rather via
a consistent truncation to N=2 supergravity in which all scalars are absent,
but
bosonic solutions of N=2 supergravity may be considered as solutions of the
field equations of (\ref{eq:twoa}) with constant $\sigma$ when $a=0$.
The values $a=1/\sqrt{3}$ and $a=\sqrt{3}$ arise from a truncation to D=5
supergravity followed by dimensional reduction to D=4. Since there is already a
vector field in the D=5 graviton supermultiplet, the resulting D=4 Lagrangian
has
{\it two} vector fields, and a further consistent truncation to one of them
yields
the above two values of $a$. As is clear from the $D=5$ origin of the
Lagrangian
(\ref{eq:twoa}) when $a=\sqrt{3}$ or $a=1/\sqrt{3}$, the scalar field $\sigma$
is
{\it not} the dilaton in this case but rather a modulus field whose expectation
value is related to the radius of the extra dimension. The value $a=1$ arises
from a
consistent truncation to N=4 supergravity, in which case $\sigma$ {\it is} the
dilaton.

The above discussion raises the following puzzle, which we now pause to
resolve.
Clearly, the Lagrangian of (\ref{eq:twoa}) with $a=1$ cannot be {\it
consistently} truncated to Maxwell-Einstein theory, but Maxwell-Einstein theory
is the bosonic sector of N=2 supergravity which {\it is} a consistent
truncation
of N=4 supergravity. There is no immediate contradiction here because
(\ref{eq:twoa}) with $a=1$ is not the bosonic Lagrangian of N=4 supergravity,
but it fails to be so only by virtue of the fact that we have discarded five of
the six vector fields. Since the same Lagrangian is found from discarding {\it
any} five of the six, it would seem that the inclusion of the other five cannot
make a difference. However, we should here recall that the vector fields of N=4
supergravity can couple to the dilaton with either $a=1$ or $a=-1$, since a
duality transformation changes the sign of $a$. As long as we have only {\it
one} vector field the sign is irrelevant but with more than one it is not. In
fact, the truncation to N=2 supergravity involves the identification of two
vector fields with opposite sign dilaton coupling constants.

Before continuing it is worthwhile to consider the limitations of the
truncation to (\ref{eq:twoa}). Although the consistency of the truncation
ensures
that any solution of the truncated field equations is a solution of the full
field equations (this is what is meant by the adjective `consistent' in this
context), the converse is not true; i.e. there will certainly be solutions of
the full untruncated equations that cannot be found among the solutions of the
truncated equations. These include dyonic black holes for $a\ne0$, which
require
a non-vanishing axion field in addition to the scalar field $\sigma$. Thus,
within the context of (\ref{eq:twoa}) we can at best verify predictions of a
$\Z_2$ subgroup of the S-duality group $Sl(2;\Z)$. However, these solutions can
be obtained \cite{STW91,KalOrt} (at least for a=1) by an S-duality
transformation of the solutions with vanishing axion field. Another class of
black hole solutions that cannot be found among solutions of the truncated
equations have been found in \cite{Cvetic}; these have also been called
`dyonic' but it should be noted that they do not carry both electric and
magnetic charge {\it of the same type}. In the context of N=8 supergravity
these
`dyonic' solutions always break more than half the supersymmetry and are
therefore not part of the Bogomol'nyi spectrum in the sense used here. To the
extent that such solutions break half the N=4 supersymmetry of the heterotic
string the discussion given here in terms of the truncated Lagrangian
(\ref{eq:twoa}) must be considered incomplete.

To make the connection between solutions of the field equations of the
truncated Lagrangian (\ref{eq:twoa}) and electrically charged particles in the
heterotic string spectrum we observe that for the truncated Lagrangian
(\ref{eq:twoa}) the Bogomolnyi bound reduces to
\begin{equation}
M^2 \ge {1\over (1+a^2)} \big[ e^{2a\langle\sigma\rangle } Q^2 +
e^{-2a\langle\sigma\rangle} P^2\big]
\label{eq:twoc}
\end{equation}
where $Q$ and $P$ are, respectively, the electric and the magnetic charge
associated with the gauge field $A$ of (\ref{eq:twoa}). Clearly, $Q$ is some
combination of the 28 electric charges and $P$ some combination of the 28
magnetic charges, the combination depending on the precise nature of the
truncation leading to (\ref{eq:twoa}).  Extreme black hole solutions saturating
the bound (\ref{eq:twoc}) exist for all values of $a$ \cite{G82,GM88,GHS91},
but
they have special properties for the particular values of (\ref{eq:twob}), some
of which will be mentioned below.

Actually, `black
hole' is an abuse of terminology since many of the solutions we shall be
concerned with are neither `black' nor `holes'. Generally we would expect
particles in the perturbative string spectrum to appear as naked singularities
(for otherwise they would have a classical structure incompatible with their
interpretation as excitations of a fundamental, i.e. structureless, string) and
we would similarly expect solutions representing particles in the
non-perturbative spectrum to appear as non-singular solitons (otherwise, there
is no obvious justification for their inclusion in the spectrum). There is no
single term that covers both these cases, not to mention several others, so
`black hole' will have to do.

If the mass formula for electric black holes, obtained by saturation of the
bound (\ref{eq:twoc}), is now compared to the mass formula of the perturbative
heterotic string spectrum for $\langle\sigma\rangle=0$ then the
result is that the $a=\sqrt{3}$ extreme electric `black holes' can be
identified
with the $q^2=0$ states and $a=1$ extreme electric black holes with the
$q^2\ge0$ states \cite{DR95}. As anticipated, these all have naked timelike
singularities. The possible roles of the $a=0$ and $a=1/\sqrt{3}$ extreme black
holes will be discussed later.

The $q^2=0$ states are the supermultiplets of Kaluza-Klein (KK) and winding
mode
states with maximum spin 2. Since these modes involve an extra dimension in an
essential way one might expect the electric $a=\sqrt{3}$ black holes to have a
D=5 interpretation. In fact, there should be two such interpretations, one for
the KK modes and the other for the winding modes. The KK interpretation arises
from the fact that the electric $a=\sqrt{3}$ black holes can interpreted as pp
waves moving in the fifth dimension \cite{GP84}; the corresponding
quanta can therefore be interpreted as particles moving at the speed of light
in the fifth dimension, but this is essentially a description of KK states. The
other interpretation is as a D=5 string. There are actually {\it
three} D=5 string solutions of the effective action of the D=5 heterotic
string, which double-dimensionally reduce to extreme black holes with
$a=1/\sqrt{3}$, $a=1$ and $a=\sqrt{3}$ \cite{LPSS95}. Obviously, the
last case is the one of relevance here.

Since the long range fields of the $q^2=0$ states are those of extreme electric
$a=\sqrt{3}$ black holes their magnetic duals must be the magnetic $a=\sqrt{3}$
black holes. These are singular in D=4 but can be interpreted either as KK
monopoles which are non-singular in D=5 \cite{S83,GP83} or
as (abelian) H-monopoles \cite{K91,GH}. The non-singularity of these solutions
is
consistent with their interpretation as solitons.  Note that the
electromagnetic $\Z_2$ subgroup of the S-duality group does {\it not} exchange
the KK states with the KK monopoles and the string winding states with the
H-monopoles, as one might expect but rather the other way around. That is, the
KK
states are exchanged with the H-monopoles and the winding modes with the KK
monopoles. The reason for this is that the KK states and the H-monopoles both
have $M\sim 1/R$, where $R$ is the radius of the fifth dimension while the
winding modes and the KK monopoles have $M\sim R$, and S-duality does not
affect
$R$. Thus, `pyrgon/KK-monopole duality' \cite{GP84,G85} in the KK sector is a
consequence of the {\it combined} S and T-dualities of the heterotic string
theory.

In the type II case, there is a simple consideration \cite{HT95a} which shows
that {\it all} charged particles with masses given by the Bogomol'nyi mass
formula, i.e. those breaking half the N=8 supersymmetry, must correspond to
$a=\sqrt{3}$ extreme black holes. The moduli space of multi-black hole
solutions
breaking half the N=8 supersymmetry must be flat, but this is the case if and
only if $a=\sqrt{3}$ \cite{R86,Sh93}. As in the heterotic case,
some of these black holes can be interpreted as fundamental string states, of
either KK or string winding type (and this identification is required by
U-duality). The remainder can all be interpreted as (non-perturbative)
p-brane `wrapping modes' \cite{HT95a}. This interpretation becomes particularly
simple in D=11. I will not go into this here as I have recently reviewed the
role of p-branes and D=11 supergravity elsewhere \cite{T95b}

It remains to consider the Bogomolnyi mass spectrum of the heterotic string
with $q^2>0$. Recall that the long range fields of these states must be those
of
$a=1$ extreme black holes. As confirmation of this identification, we
note that, since a charge vector with positive $SO(6,22)$ norm can be rotated
to
one that is carried only by vector fields in the graviton supermultiplet, this
sector of the Bogomol'nyi spectrum is essentially of purely gravitational
origin. This leads one to expect these particles to appear as extreme
black hole solutions of N=4 supergravity, for which the `dilaton' coupling
constant is indeed $a=1$. Moduli space considerations provide a check on this
because the moduli space of solutions of an $N=4$ supersymmetric field theory
which break half the supersymmetry must be hyper-K{\"a}hler \cite{Gauntlett}.
At
first sight, this condition appears not to be fulfilled because the moduli
space
of $a=1$ black hole solutions of the field equations of (\ref{eq:twoa}) has the
wrong dimension to be hyper-K{\"a}hler, but it must be remembered that the
Lagrangian (\ref{eq:twoa}) is {\it not} the bosonic sector of an N=4
supergravity
theory, even for $a=1$, but rather a consistent truncation of one. It follows
that the moduli space need not be hyper-K{\"a}hler but must be a totally
geodesic submanifold of a hyper-K{\"a}hler manifold, as indeed it is
\cite{GM95}.

One might wonder why an $a=1$ black hole cannot decay into two
$a=\sqrt{3}$ black holes given that, according to the mass formula
(\ref{eq:twoc}), the mass of an $a=1$ black hole with charge $Q$ is greater
than the sum of any number of $a=\sqrt{3}$ extreme black holes with total
charge $Q$. This is the wrong comparison, however, since the charge carried by
an $a=1$ black hole cannot be identified with the charge carried by the
$a=\sqrt{3}$ black holes. The point is that $a=\sqrt{3}$ extreme electric black
holes have $q^2=0$, so that they must carry more than one charge of the
untruncated theory and, consequently, the addition of charges is {\it vector}
addition. Consider the decay of an $a=1$ extreme black hole of charge
$Q_{(1)}$ into two $a=\sqrt{3}$ extreme black holes, each of charge
$Q_{(\sqrt{3})}$; in order that the charge vectors of the latter sum to the
charge vector of the former they must be orthogonal, so charge conservation
implies that $Q_{(1)}= \sqrt{2}Q_{(\sqrt{3})}$, rather than
$Q_{(1)}= 2Q_{(\sqrt{3})}$. Application of the mass formula (\ref{eq:twoc}) now
shows that although the decay is still energetically possible it has no
available phase space and so cannot proceed.

As noted earlier, in the $a=1$ case the scalar $\sigma$ in (\ref{eq:twoa}) is
the dilaton so $g= e^{\langle\sigma\rangle}$ is the string coupling constant
and we can therefore rewrite the mass formula (\ref{eq:twoc}) as
\begin{equation}
M^2 = {1\over2} \big[ g^2 Q^2 + g^{-2} P^2\big]\ .
\label{eq:threeb}
\end{equation}
But this cannot be right in the string theory context because electrically
charged particles would all have zero mass at $g=0$. The resolution
\cite{W95,HPC} involves the realization that the Lagrangian from which
(\ref{eq:threeb}) was derived was expressed in terms of the canonical, or `
Einstein', metric $ds^2$, whereas we should be using the {\it string metric}
$d\tilde s^2= g^2 ds^2$. A simple way to obtain the formula appropriate to the
string metric is to observe that the ADM mass $M$ of an asymptotically-flat
spacetime depends implicitly on the normalization at spatial infinity,
$\lim(-k^2)$, of an asymptotic timelike Killing vector field $k$. To make this
dependence explicit we must replace (\ref{eq:threeb}) by \begin{equation}
M^2 = {1\over2} \big[ g^2 Q^2 + g^{-2} P^2\big]\lim(-k^2)\ .
\label{eq:threec}
\end{equation}
Normally we choose $\lim(-k^2)=1$ where $k^2$ is defined in terms of the
`Einstein' metric, but in the string theory context we should instead define
$\lim(-\tilde k^2)=1$ where $\tilde k^2$ is defined in terms of the string
metric. Since $\tilde k^2 =g^2 k^2$ we should set $\lim(-k^2) = g^{-2}$ in
(\ref{eq:threec}) to get
\begin{equation}
M^2 = Q^2 + {1\over g^4}P^2\ ,
\label{eq:threed}
\end{equation}
where the electrically charged particles now have masses that are independent
of $g$, as is appropriate for particles in the perturbative string spectrum,
while the magnetically charged particles have masses with the
$g^{-2}$ dependence expected of solitons.

In the type II case the formula analogous to (\ref{eq:threed}) is , {\it
schematically},
\begin{equation}
\label{eq:oneh}
M^2 = q^2 + {1\over g^4}p^2 + {1\over g^2} r^2\ ,
\end{equation}
where $q$ and $p$ now represent, respectively, the NS-NS electric and
magnetic charges and $r$ represents a RR charge, either electric or magnetic.
As the formula shows, both magnetic {\it and} electric RR charges can appear
only non-perturbatively as required by U-duality \cite{HT95a}, but their masses
are larger than is usual for solitons by  a factor of $g^{-1}$ \cite{W95}. As
mentioned earlier, only the $a=\sqrt{3}$ black holes saturate the Bogomol'nyi
bound in the type II case because only these preserve half the supersymmetry.
The $a=0$ black holes are, of course, also solutions of N=8 supergravity,
because they were solutions for N=4, but in the N=8 context they preserve only
$1/4$ of the supersymmetry. Presumably they will therefore be unstable against
decay into $a=\sqrt{3}$ black holes.  The same argument can be used to dispose
of the $a=0$ and $a=1/\sqrt{3}$ black holes in the N=4 case since both preserve
only $1/4$ of the N=4 supersymmetry. The values of $a$ for which the extreme
black holes saturate the N-extended Bogomol'nyi bound is shown in the table
below.

\begin{table}[hbt]
\caption{Bogomol'nyi Black Holes}
\small
\begin{tabular}{||c|c||}
\hline\hline
No. of Susys & Scalar Coupling \\
\hline\hline
N=8 & $a^2=3$ \\
\hline
N=4 & $a^2=1,3$ \\
\hline
N=2 & $a^2=0,{1\over3},1,3$ \\
\hline\hline
\end{tabular}
\end{table}
Thus, as long as we are concerned with stable
particles of $N=4$ and $N=8$ supestring theories we need not consider the
$a=0$ and $a=1\sqrt{3}$ cases. Nevertheless, it is interesting to consider
what their study might tell us about N=2 superstring compactifications,
although one should bear in mind that renormalization effects might change the
picture.

Consider first the $a=0$ extreme black holes. The first point to note
is that the electric and magnetic cases have an {\it identical} metric, the
extreme Reissner-Nordstrom metric. So, on the evidence of supergravity
solutions, both should be given the {\it same} interpretation, either
fundamental or solitonic. It is impossible for both to be (simultaneously)
fundamental so a solitonic interpretation of both the magnetic {\it and} the
electric $a=0$ black holes is the only option. This would make
them similar to RR solitons in type II superstring theories.
However, the RN metric has a timelike singularity, admittedly hidden behind
an event horizon, so the soliton interpretation is problematic and there
there is therefore no really compelling reason to include $a=0$ black holes as
part of a superstring spectrum. It is instructive to compare this situation
with
that of $a=1$ extreme black holes. The Einstein metric is again the same for
both the electric and magnetic cases but the relevant metric is now the string
metric, in terms of which the electric and magnetic solutions differ radically:
the electric $a=1$ black hole has a naked timelike singularity, in accord with
its fundamental status, while the magnetic $a=1$ black hole is geodesically
complete \cite{GHS91}, in accord with its solitonic status.

Since the Lagrangian (\ref{eq:twoa}) has a D=5 interpretation for
$a=1/\sqrt{3}$, one should not be surprised to discover that the $a=1/\sqrt{3}$
black holes have a D=5 interpretation. Indeed, the electric extreme
$a=1/\sqrt{3}$ black hole is just a dimensional reduction of the D=5 RN black
hole \cite{MP86,M87} while the magnetic extreme
$a=1/\sqrt{3}$ black hole is interpretable as a double dimensional reduction of
a D=5 string \cite{GHT95}. The global structure of these solutions is more
promising than for the $a=0$ case because although both the D=5 black hole and
the D=5 string have event horizons, the string is nevertheless geodesically
complete. The situation here is entirely analogous to that of the electric
membrane and magnetic fivebrane of D=11 supergravity \cite{DGT94,GHT95}, which
arise in a similar manner from the D=10 type IIA electric membrane and magnetic
fourbrane. This is but one aspect of the quite close analogy between D=5 and
D=11 that may well repay future study.

\end{document}